\documentclass{ws-ijmpe}
\usepackage[super,compress]{cite}
\usepackage{subcaption}  
\usepackage{comment} 
\usepackage{epstopdf}  %
\usepackage{float}  
\usepackage{hyperref}
\hypersetup{
	colorlinks   = true, 
	urlcolor     = blue, 
	linkcolor    = blue, 
	citecolor   = blue 
}
\usepackage{makecell}
\begin{document}

\markboth{Y. EL BASSEM $\&$ M. OULNE}{NUCLEAR STRUCTURE INVESTIGATION OF EVEN-EVEN AND ODD Pb  ....}

\catchline{}{}{}{}{}

\title{NUCLEAR STRUCTURE INVESTIGATION OF EVEN-EVEN AND ODD Pb ISOTOPES BY USING THE HARTREE-FOCK-BOGOLIUBOV METHOD}

\author{Y. EL BASSEM\footnote{corresponding author.} ~and M. OULNE$^\dagger$}

\address{High Energy Physics and Astrophysics Laboratory, Department of Physics, \\Faculty of Sciences SEMLALIA, Cadi Ayyad University,  \\P.O.B. 2390, Marrakesh, Morocco.\\
$^*$younes.elbassem@ced.uca.ma\\
$^\dagger$oulne@uca.ma}

\maketitle

\begin{history}
\received{Day Month Year}
\revised{Day Month Year}
\end{history}

\begin{abstract}
The nuclear structure of even-even and odd lead isotopes ($^{178-236}$Pb) is investigated within the Hartree-Fock-Bogoliubov theory. Calculations are
performed for a wide range of neutron numbers, starting from the proton-rich side up to the neutron-rich side, by using the SLy4 Skyrme interaction and a new proposed formula for the pairing strength which is more precise for this region of nuclei as we did in previous works in the regions of Neodymium (Nd, Z=60) [\href{http://www.worldscientific.com/doi/abs/10.1142/S0218301315500731}{Int. J. Mod. Phys. E 24, 1550073 (2015)}] and Molybdenum (Mo, Z=42) [\href{http://www.sciencedirect.com/science/article/pii/S0375947416301762}{Nuc. 
	Phys. A 957 22-32 (2017)}]. Such a new pairing strength formula allows reaching exotic nuclei region where the experimental data are not available. Calculated values of various physical quantities such as binding energy, two-neutron separation energy, quadrupole deformation, and rms-radii for protons and neutrons are discussed and compared with experimental data and some estimates of other nuclear models like Finite Range Droplet Model (FRDM), Relativistic Mean Field (RMF) model with NL3 functional (NL3), Density-Dependent Meson-Exchange Relativistic Energy Functional (DD-ME2) and results of Hartree-Fock-Bogoliubov calculations based on the D1S Gogny effective nucleon-nucleon interaction (Gogny D1S).
\end{abstract}

\keywords{Hartree-Fock-Bogoliubov method; $Pb$ isotopes; binding energy; proton, neutron and charge radii; quadrupole deformation.}

\ccode{PACS numbers: }


\section{Introduction}

\label{intro}
A long-standing goal of research in nuclear physics is to make reliable predictions with one nuclear model in order to describe the ground-state properties of all nuclei in the nucleic chart. For this purpose, several approaches have been developed to study ground-state and single-particle (s.p) excited states properties of nuclei.
For light nuclei with a mass number up to $A\approx12$, \textit{ab initio} calculations employing the realistic nucleon-nucleon and  three-nucleon interactions with the Green's function Monte Carlo (GFMC) method are applicable  \cite{Pudliner1997,Pudliner1995}. Another line of approach is the No-Core Shell Model (NCSM) of Barett's group \cite{Navratil1997,Navratil1998}, where all nucleons  are treated in a large number of shell-model basis, providing similarly successful description of light nuclei. 
While these methods have been successful in the case of light nuclei, their applications to heavier systems become rather difficult due to the numerical difficulties in handling the nuclear many-body problem.
To overcome this difficulty, many approaches have been introduced. Among them, the Shell Model Monte Carlo (SMMC) method which has been successfully proposed \cite{Koonin} and is basically suitable for the ground state and thermal properties.
A completely different approach, the Monte Carlo Shell Model (MCSM), has been introduced \cite{Otsuka1999,Otsuka2001} for nuclear shell model calculations \cite{Covello2010}.
For medium-mass nuclei up to $A\approx60$, the large-scale shell model \cite{Caurier2002,Brown1988} may be used. While for heavier nuclei, non-relativistic \cite{Vautherin1972,Decharge1980,Terasaki,Dobaczewski1996,Chabanat1998,Stoitsov2000,Teran} and relativistic \cite{Ring96,Lalazissis,Meng2006,Long2004,Zhao2010} mean field theories have received much attention for describing the ground-state properties of nuclei. The Relativistic Mean Field (RMF) theory with a limited number of parameters can describe a large amount of data such as: the pseudo-spin symmetry \cite{Meng1999,Ti-Sheng2003,Zhou2003}, the chiral symmetry \cite{Meng2006_2}, and the spin-isospin excitations \cite{Liang2009,Liang2008}. In spite of these successes, there are still a number of questions needed to be answered in the RMF theory: the contributions due to the exchange (Fock) terms and the pseudo-vector $\pi$-meson. There exist attempts to include the exchange terms in the relativistic description of nuclear matter and finite nuclei. For instance:
Considerable effort has been devoted to relativistic Hartree-Fock (RHF) theory \cite{Long2006,Long2007,Long2010}.
However, because of its numerical complexity, for a long
time, it failed in a quantitative description of nuclear systems.
 One of the most important phenomenological approaches widely used in nuclear structure calculations is the Hartree-Fock-Bogoliubov method \cite{Bogoliubov1959}, which allows particles and holes to be treated on an equal footing by unifying the self-consistent description of nuclear orbitals, as given by Hartree-Fock (HF) approach, and the Bardeen-Cooper-Schrieffer (BCS) pairing theory \cite{Bardeen} into a single variational method.

In our previous works, Refs. \refcite{ElBassemYounes2015} and \refcite{ElBassemYounes2017}, we have investigated the ground-state properties within the Hartree-Fock-Bogoliubov method with SLy5 and SLy4 skyrme forces in the regions of Nd and Mo, respectively. In each case the pairing strength has been generalized with a new proposed formula. The results obtained in those works (Refs. \refcite{ElBassemYounes2015} and \refcite{ElBassemYounes2017}), including the binding energy per nucleon, the one- and two-neutron separation energies, proton, neutron and charge radii and quadrupole deformation, were in a good agreement with the available experimental data in comparison with other theoretical models such as the Relativistic Mean Field model with NL3 functional (NL3), the Density-Dependent Meson-Exchange Relativistic Energy Functional (DD-ME2) and results of Hartree-Fock-Bogoliubov calculations based on the D1S Gogny effective nucleon-nucleon interaction (Gogny D1S).
In this work, we will extend our investigation to another region which is that of the lead (Pb, $Z=82$). 

The region of lead (Pb) is very important for studying the nuclear structure, since $Z=82$ is a typical proton magic number. Plenty of experimental data and many theoretical models have been employed to investigate the nuclei in this region \cite{Sharma1993,Anselment1986,Egido2004,Rodrguez2004}. In the present work, we intend to check the validity of our treatment on the structure and properties of lead isotopes. In this treatment we will use the HFB method with SLy4 Skyrme interaction \cite{Chabanat1998} and a new proposed formula for the pairing strength.

The paper is organized as follows. In Section \ref{section2}, a brief review of the Hartree-Fock-Bogoliubov method is provided as well as details about the numerical calculations. While Section \ref{section3} is devoted to present our results and discussion. A summary of the present study is given in Section \ref{section4}.
\section{Theoretical framework}
\label{section2}
\subsection{Hartree-Fock-Bogoliubov Method}

The Hartree-Fock-Bogoliubov (HFB)\cite{Bogoliubov1959,Ring}  method with effective zero-range pairing forces is a reliable and computational convenient way to study the nuclear pairing correlations in both stable and unstable nuclei.
The HFB framework has been extensively discussed in the literature \cite{Ring,Dobaczewski84,Dobaczewski96,Bender} and will be briefly introduced here.\\
In the standard HFB method, a two-body Hamiltonian of a system of fermions can be expressed in terms of a set of annihilation and creation operators $(c, c^\dagger)$:

\begin{equation}
H=\sum_{n_1 n_2} e_{n_1 n_2} c_{n_1}^\dagger c_{n_2} + \frac{1}{4} \sum_{n_1 n_2 n_3 n_4} \bar{\nu}_{n_1 n_2 n_3 n_4} c_{n_1}^\dagger c_{n_2}^\dagger c_{n_4} c_{n_3}
\label{eq1}
\end{equation}
with the first term corresponding to the kinetic energy and $\bar{\nu}_{n_1 n_2 n_3 n_4}=\langle n_1 n_2 | V | n_3 n_4 - n_4 n_3 \rangle$ are anti-symmetrized two-body interaction matrix-elements. \vspace{0.4em}
In HFB method, the ground-state wave function $|\varPhi\rangle$ is defined as the quasi-particle vacuum $\alpha_k|\varPhi\rangle=0$, in which the quasi-particle operators $(\alpha,\alpha^\dagger)$ are connected to the original particle ones via a linear Bogoliubov transformation:
\begin{equation}
\alpha_k=\sum_n (U_{nk}^* c_n + V_{nk}^* c_n^\dagger),~~~~~~~~\alpha_k^\dagger=\sum_n (V_{nk} c_n + U_{nk} c_n^\dagger),
\end{equation}
which can be rewritten in the matrix form as:
\begin{eqnarray}
\left(\begin{array}{c} \alpha \\ \alpha^\dagger \end{array}\right)=\left(\begin{array}{c  c} U^\dagger & V^\dagger \\
V^T & U^T
\end{array} \right)\left(\begin{array}{c} c \\ c^\dagger \end{array} \right) , \, \label{eqhfb}
\end{eqnarray}
The matrices U and V satisfy the relations:
\begin{eqnarray}
\begin{split}
&U^\dagger U+V^\dagger V=1,~~~~~~~U U^\dagger + V^* V^T=1,\\
&U^T V+V^T U =0,~~~~~~U V^\dagger+ V^* U^T=0.
\end{split}
\end{eqnarray}
The basic building blocks of this theory are the density matrix and the pairing tensor. In terms of the normal $\rho$ and pairing $\kappa$ one-body density matrices, defined as:
\begin{eqnarray}
\begin{split}
&\rho_{nn'}=\langle\Phi|c_{n'}^\dagger c_{n}|\Phi\rangle=(V^{*}V^{T})_{nn'}, \hspace{0.3 in}\\
&\kappa_{nn'}=\langle\Phi|c_{n'}c_{n}|\Phi\rangle=(V^{*}U^{T})_{nn'}\,,
\label{matrices}
\end{split}
\end{eqnarray}
the expectation value of the Hamiltonian (\ref{eq1}) is expressed as an energy functional
\begin{equation}
E[\rho,\kappa]=\frac{\langle\Phi|H|\Phi\rangle}{\langle\Phi|\Phi\rangle}=
\textrm{Tr}[(e+\frac{1}{2}\Gamma)\rho]-\frac{1}{2}\textrm{Tr}[\Delta\kappa^{*}]
\label{energyfunctional}
\end{equation}
where\\
\begin{eqnarray}
\begin{split}
&\Gamma_{n_{1}n_{3}}=\sum_{n_{2}n{4}}\bar\upsilon_{n_{1}n_{2}n_{3}n_{4}}\rho_{n_{4}n_{2}} \\         &\Delta_{n_{1}n_{2}}=\frac{1}{2}\sum_{n_{3}n{4}}\bar\upsilon_{n_{1}n_{2}n_{3}n_{4}}\kappa_{n_{3}n_{4}}\,.
\end{split}
\end{eqnarray}
The variation of the energy (\ref{energyfunctional}) with respect to $\rho$ and $\kappa$ leads to the HFB equations:
\begin{eqnarray}
\left(\begin{array}{cc} e+\Gamma-\lambda & \Delta \\
-\Delta^* & -(e+\Gamma)^*+\lambda
\end{array} \right)\left(\begin{array}{c} U \\ V \end{array} \right) =E\left(\begin{array}{c} U \\ V \end{array}\right), \, \label{eqhfb}
\end{eqnarray}
where $\Delta$ and $\lambda$ denote the pairing potential and Lagrange multiplier, introduced to fix the correct average particle number, respectively.\\
It should be stressed that the energy functional (\ref{energyfunctional}) contains terms that cannot be simply related to some prescribed effective interaction \cite{Bender}. 
In terms of Skyrme forces, the HFB energy (\ref{energyfunctional})
has the form of local energy density functional:

\begin{equation}
E[\rho,\tilde{\rho}]=\int d^{3}\textbf{r}\textrm{H}(\textbf{r}),
\label{skyrmeefunctional}
\end{equation}
where \textrm{H}(\textbf{r})
is the sum of the mean field and pairing energy densities. 
\begin{equation}
\textrm{H}(\textbf{r})=H(\textbf{r})+\tilde{H}(\textbf{r})
\end{equation}
We use the pairing density matrix $\tilde{\rho}$,
\begin{equation}
\tilde{\rho}(\textbf{r}\sigma,\textbf{r}^{\prime}\sigma^{\prime})=-2 \sigma^{\prime} \kappa(\textbf{r},\sigma,\textbf{r}^{\prime},-\sigma^{\prime})
\end{equation}
instead of the pairing tensor $\kappa$. This is convenient for describing time-even quasiparticle states when both $\rho$ and $\tilde{\rho}$ are hermitian and time-even \cite{Dobaczewski84}.\\
The variation of the energy (\ref{skyrmeefunctional}) with respect to the particle local density $\rho$ and pairing local density $\tilde{\rho}$ results in the Skyrme HFB equations:

\begin{eqnarray}
\begin{split}
\sum_{\sigma^{\prime}}\left(\begin{array}{cc} h(\textbf{r},\sigma,\sigma^{\prime}) & \Delta(\textbf{r},\sigma,\sigma^{\prime}) \\
\Delta(\textbf{r},\sigma,\sigma^{\prime}) & -h(\textbf{r},\sigma,\sigma^{\prime})
\end{array} \right)\left(\begin{array}{c} U(E,\textbf{r}\sigma^{\prime}) \\
V(E,\textbf{r}\sigma^{\prime})\end{array} \right) =\\
\left(\begin{array}{cc} E+\lambda & 0\\
0 & E-\lambda\end{array}\right)\left(\begin{array}{c} U(E,\textbf{r}\sigma) \\
V(E,\textbf{r}\sigma) \end{array}\right) \,\,
\label{shfb}
\end{split}
\end{eqnarray}
where $\lambda$ is the chemical potential. The local fields $h(\textbf{r},\sigma,\sigma^{\prime})$
and $\Delta(\textbf{r},\sigma,\sigma^{\prime})$ can be calculated in coordinate space. Details can be found in Refs.~[\refcite{Ring,Stoitsov2005,Greiner}].

\subsection{Details of Calculations}
In this work, the ground-state properties of even-even and odd $^{178-236}Pb$ have been reproduced by using the computer code HFBTHO (v2.00d) \cite{Stoitsov2013} which utilizes the axial Transformed Harmonic Oscillator (THO) single-particle basis to expand quasi-particle wave functions. It iteratively diagonalizes the Hartree-Fock-Bogoliubov Hamiltonian based on generalized Skyrme-like energy densities and zero-range pairing interactions until a self-consistent solution is found.

Calculations were performed with the SLy4 Skyrme functional \cite{Chabanat1998} as in Ref.~\refcite{Stoitsov2005}, and by using the same parameters as in our previous works \cite{ElBassemYounes2015,ElBassemYounes2017}: A quasi-particle cutoff of $E_{cut}=60~MeV$, which means that only quasi-particle states with energy lower than $60~MeV$ are taken into account, and a mixed surface-volume pairing with identical pairing strength for both protons and neutrons. 
The number of oscillator shells taken into account was $N_{max}=20~shells$, the total number of states in the basis $N_{states}=500$, and the value of the deformation $\beta$ is taken from the column $\beta_2$ of the Ref.~\refcite{Moller1995}. The THO basis is characterized by the frequency $\hbar \omega_0=1.2*41/A^{1/3}$, 
and the  length $b_0$ is automatically sets by using the relation $b_0=\sqrt{\hbar / m \omega_0}$. 
The number of Gauss-Laguerre and Gauss-Hermite quadrature points was $N_{GL} = N_{GH} = 40$, and the number of Gauss-Legendre points for the integration of the Coulomb potential was $N_{Leg} = 80$ \cite{ElBassemYounes2015,ElBassemYounes2017}.

Odd-N isotopes are calculated by using the blocking of quasi-particle states \cite{Schunck}. The time-reversal symmetry is, by construction, conserved in the computer code HFBTHO (v2.00d), the blocking prescription is implemented in the equal filling approximation \cite{Schunck,Perez}, and the time-odd fields of the Skyrme functional are identically zero. The identification of the blocking candidate is done using the same technique as in HFODD \cite{Dobaczewski2009}: the mean-field Hamiltonian $h$ is diagonalized at each iteration and provides a set of equivalent single-particle states. Based on the Nilsson quantum numbers of the requested blocked level provided in the input file, the code identifies the index of the quasi-particle (q.p.) to be blocked by looking at the overlap between the q.p. wave-function (both lower and upper component separately) and the s.p. wave-function. The maximum overlap specifies the index of the blocked q.p. \cite{Stoitsov2013}.

In the present work, among several parameters sets for prediction of the nuclear ground state properties \cite{Bartel,Baran}, we used the SLy4 Skyrme force \cite{Chabanat1998} which is widely used in nuclear structure calculations. 
The SLy4 Skyrme force was developed by E. Chabanat and his collaborators. It is part of the SLyx family of forces. To determine the set of parameters (10 in total), the authors followed an adjustment protocol including certain properties of the infinite nuclear matter (saturation value $\rho_0$ and incompressibility coefficient) and some properties of the finite nuclear matter (masses and radii of some doubly magic nuclei). This force was built for extreme conditions of isospin and density. Nevertheless, it leads to very good results for nuclei of the valley of stability and strongly deformed nuclei. 
SLy4 \cite{Chabanat1998} parameters set used in this study are shown in Table~\ref{table1}.

\begin{table}[ht]
	\centering
	\caption{SLy4 parameters set considered in this work.}
	\label{table1}
	\begin{tabular}{ll} 
		\hline\noalign{\smallskip}
		Parameter & SLy4 \\
		\noalign{\smallskip}\hline\noalign{\smallskip}
		t$_0$ (MeV fm$^3$)   &     -2488.91 \\
		t$_1$ (MeV fm$^5$)   &       486.82 \\
		t$_2$ (MeV fm$^5$)   &      -546.39 \\
		t$_3$ (MeV fm$^4$)   &       13777.0 \\
		x$_0$ 		    &       0.834 \\
		x$_1$    	    &       -0.344 \\
		x$_2$   	    &      -1.000 \\
		x$_3$  		    &       1.354 \\
		W$_0$ (MeV fm$^3$)   &       123.0	\\
		$\sigma$            &        1/6    \\ 
		\noalign{\smallskip}\hline
	\end{tabular}
\end{table}

As in Refs. \refcite{ElBassemYounes2015} and \refcite{ElBassemYounes2017}, in the input data file of the computer code HFBTHO (v2.00d) \cite{Stoitsov2013}, we have modified the values of the pairing strength for neutrons $V_0^{n}$ and protons $V_0^{p}$ (in MeV), which may be different, but in our study the pairing strength $V_0^{n,p}$ is taken to be the same for both. For each isotope, we have executed the code by using a given $V_0^{n,p}$ and compared the obtained ground-state total binding energy with the experimental value. This procedure was repeated until we found, for each mass number A (both odd and even), the value of $V_0^{n,p}$ that gives the ground-state total binding energy closest to the experimental one. For more details, see Refs. [\refcite{ElBassemYounes2015,ElBassemYounes2017}] and references therein.

By fitting the obtained values of $V_0^{n,p}$ to the mass number $A$, we have found the following formula:
\begin{equation}
\large {{ V_0^{n,p} = 7.28\,A^{\frac{3}{4}}}}
\label{eqV0}
\end{equation}
The equation (\ref{eqV0}) gives, for each isotope of mass number A (both even and odd), the pairing-strength $V_0^{n,p}$ appropriate to reproduce the binding energies closest to the experimental ones of lead isotopes.

In order to check the validity of Eq.~(\ref{eqV0}), it has been used to generate the pairing-strength $V_0^{n,p}$ for each mass number $A$, then we have included this value of $V_0^{n,p}$ in the input data file of the computer code HFBTHO (v2.00d) so as to calculate several ground-state properties such as binding energy, two-neutron separation energy, neutron and proton radii and quadrupole deformation. 
The obtained results are presented in the next section.

\section{Results and Discussion}
\label{section3}
In this section we present the numerical results of this work, particularly for binding energy, two-neutron separation energy, charge, neutron and proton radii and quadrupole deformation for $^{178-236}$Pb isotopes.\\
In all our calculations, we used the Skyrme force (SLy4) and Eq.~(\ref{eqV0}) for the pairing strength.

\subsection{Binding energy}

\begin{figure}[!htb]
	\centerline{\psfig{file=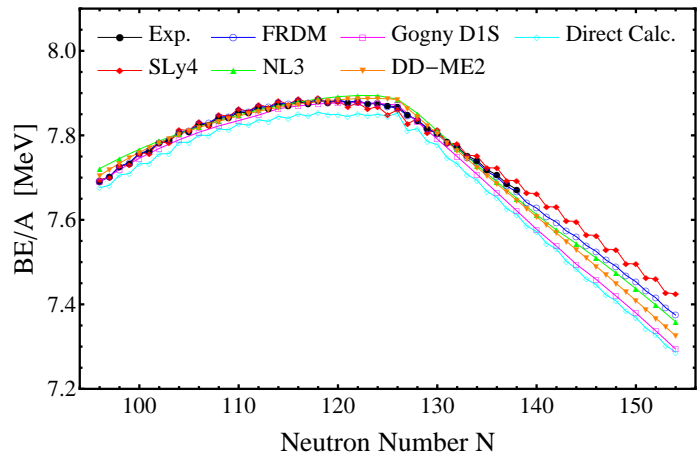,height=6cm}}
	\caption{
		(Color online) The calculated binding energies per nucleon for $Pb$ isotopes obtained from the HFB calculations with Sly4, in comparison with experimental data and those obtained by FRDM, Gogny D1S, NL3,
		DD-ME2, and Direct Calculation.}
	\label{BEexp}
\end{figure}

Binding Energy (BE) is a very important quantity in nuclear physics and one of the key observables for understanding the structure of a nucleus. 
The calculated Binding Energies per nucleon (BE/A) for lead isotopes, obtained by using the pairing strength generated by Eq.~(\ref{eqV0}) are plotted in Fig.~\ref{BEexp}, as function of the neutron number $N$, and compared with those obtained by direct calculation ("Direct Calc.") which means by using the default value of the pairing strength which is a pre-defined pairing force used in the computer code for each Skyrme functional regardless of the mass number, in the case of SLy4, the default values are : $V_0^{n} =-325.2500$ and $V_0^{p} =-340.0625$ for neutrons and protons, respectively. 
The experimental binding energies per nucleon \cite{WANG} as well as the HFB calculations based on the D1S Gogny force \cite{AMEDEE}, the predictions of the Finite Range Droplet Model (FRDM) \cite{Moller97}, the Relativistic Mean Field (RMF) model with NL3 functional \cite{Lalazissis} and the Density-Dependent Meson-Exchange Relativistic Energy Functional (DD-ME2) \cite{Lalazissis2005} are also shown in Fig.~\ref{BEexp} for comparison.

From Fig.~\ref{BEexp}, we note that the maximum in the binding energy per nuclei, (BE/A), for lead isotopes is observed at the magic neutron number $N = 126$.

\begin{figure}[!htb]
	\minipage{0.48\textwidth}
	\centerline{\psfig{file=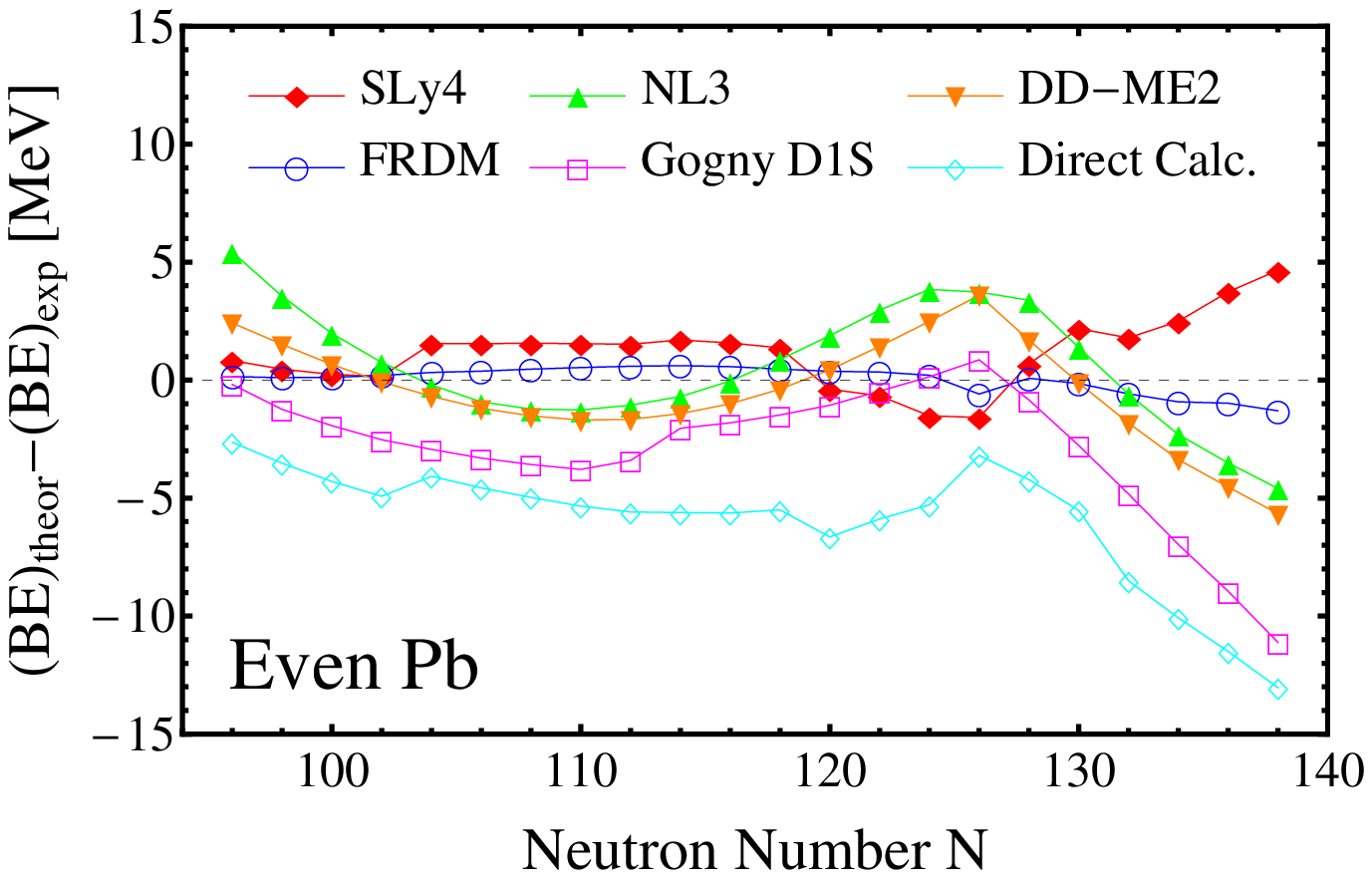,width=\linewidth, height=4.5cm}}
	\endminipage\hfill
	\minipage{0.48\textwidth}
	\centerline{\psfig{file=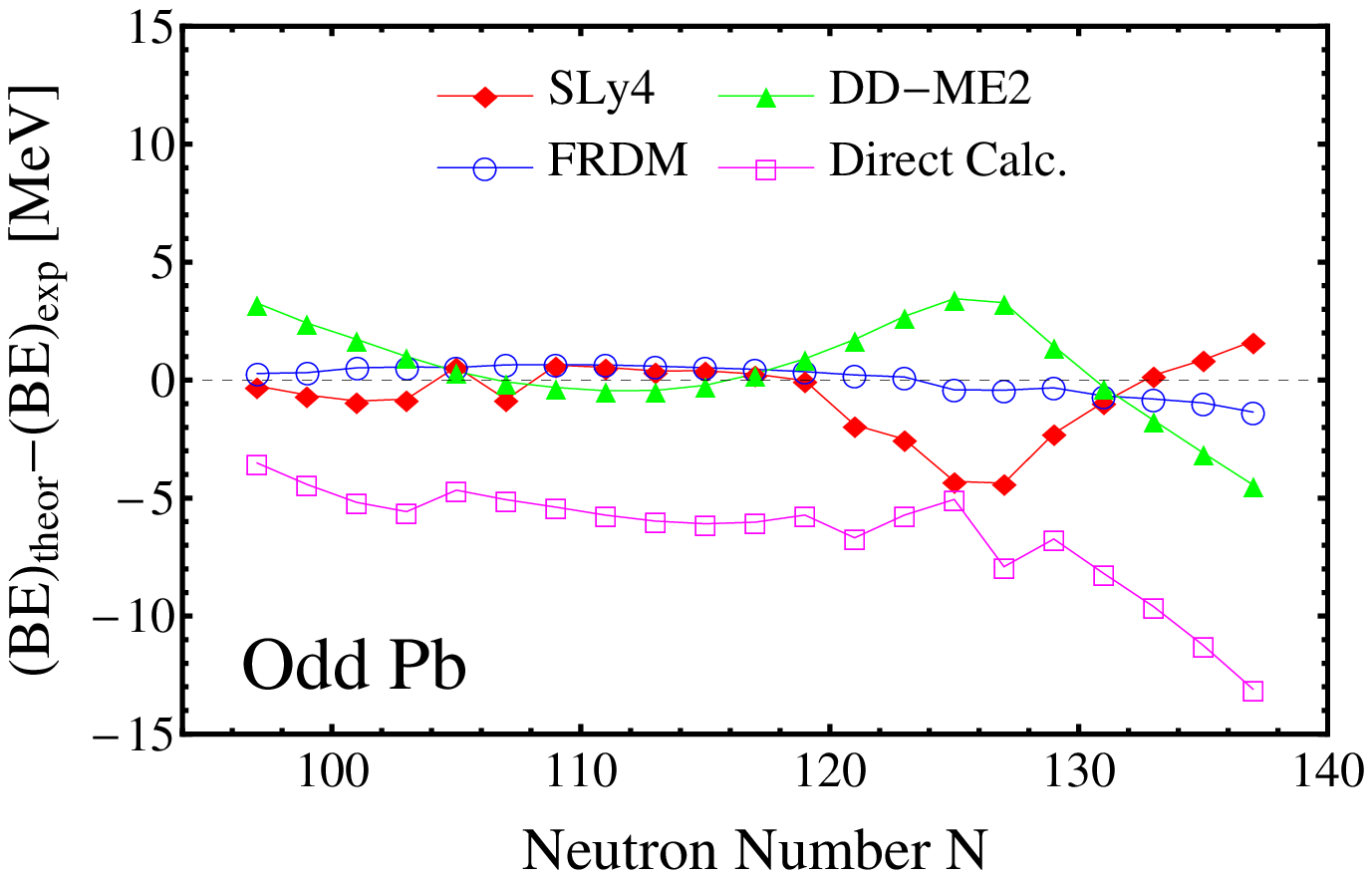,width=\linewidth, height=4.5cm}}
	\endminipage\hfill
	\caption{(Color online) Differences between calculated results of total binding energies and experimental values for even isotopes (left panel) and odd ones (right panel), in comparison with data obtained by other available theoretical models.}
	\label{Delta_BE}
\end{figure}

In order to provide a further check of the accuracy of our results, the differences between the experimental total BE and the calculated results obtained in this work by using Eq.~(\ref{eqV0}) are shown as function of the neutron number $N$ in Fig.~\ref{Delta_BE} for even isotopes (left panel) and for odd ones (right panel), separately. The results of other nuclear models are also included for comparison. 
We point out that this comparison is made only for isotopes that have available experimental data.

The root-mean-square (rms) deviation between the calculated results in the
present study and the experimental data are listed in Tables~\ref{table_BE_even} and \ref{table_BE_odd} for even and odd isotopes, respectively. The predictions of FRDM \cite{Moller97} and NL3 \cite{Lalazissis} theories as well as the HFB calculations based on the D1S Gogny force \cite{AMEDEE} are listed too for comparison.

\begin{table}[!htb]
	\caption{The root mean square deviations of the calculated binding energies of the even $Pb$ isotopes from the experimental data, in comparison with other theoretical models.
		\label{table_BE_even}}
	\centering
	\begin{tabular}{llllll} 
		\hline\noalign{\smallskip}
		SLy4 	&   Direct Calc.	&	NL3		&  FRDM		&	Gogny D1S &   DD-ME2 \\ 
		\noalign{\smallskip}\hline\noalign{\smallskip}
		1.8564 	&	6.4609	&	2.5795	&  0.5482 	&	4.0746	  &   2.2670 \\  
		\noalign{\smallskip}\hline
	\end{tabular}
\end{table}

\begin{table}[!htb]
	\caption{Same as Table~\ref{table_BE_even}, but for odd isotopes.\label{table_BE_odd}}
	\centering
	\begin{tabular}{llllll} 
		\hline\noalign{\smallskip}
		SLy4 	&   Direct Calc.	&  FRDM		&	   DD-ME2 \\ 
		\noalign{\smallskip}\hline\noalign{\smallskip}
		1.7023 	&	6.9293	&	0.6035	&  2.0616 	 \\  
		\noalign{\smallskip}\hline
	\end{tabular}
\end{table}

As we can see from Tables~\ref{table_BE_even} and \ref{table_BE_odd}, in both cases odd and even lead isotopes, our work exceeds in accuracy all the other nuclear models, except the FRDM which is the most accurate. In addition, the use of our proposed pairing strength formula (Eq.~\ref{eqV0}) has clearly improved the accuracy of the results in comparison with those obtained in direct calculations.

\subsection{Neutron separation energy}
\begin{figure}[!htb]
	\centerline{\psfig{file=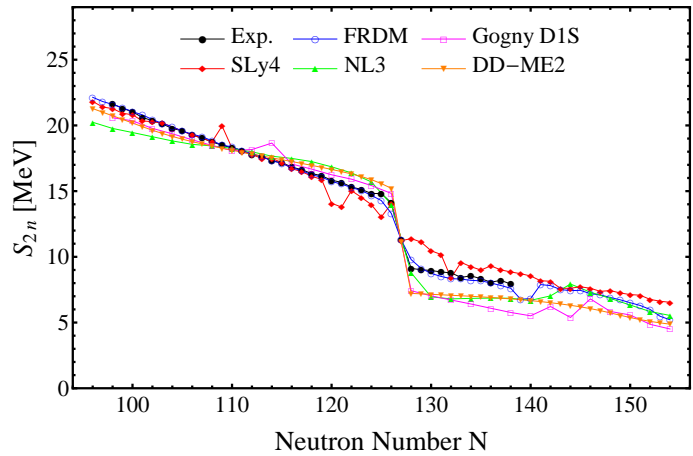,height=6cm}}
	\caption{(Color online) The calculated double neutron separation energies, $S_{2n}$, for $Pb$ isotopes obtained from the HFB calculations with Sly4, in comparison with experimental data and those obtained by FRDM, Gogny D1S, NL3, and DD-ME2. }
	\label{S2n}
\end{figure}

The two-neutron separation energy, $S_{2n}$, is an important quantity in investigating the nuclear shell structure and in testing the validity of a model. 
In the present work, we calculated the two-neutron separation energies for lead isotopes in SLy4 parametrization with the pairing strength $V_0^{n,p}$ generated by Eq.~(\ref{eqV0}).
The double neutron separation energy is defined as:
\begin{equation}
S_{2n}(Z,N)=BE(Z,N)-BE(Z,N-2)
\end{equation}
Note that in using this equation, all binding energies must be entered with a positive sign. The calculated $S_{2n}$ for lead isotopes are displayed in Fig.~\ref{S2n}. The available experimental data \cite{WANG}, and the predictions of other nuclear models such as NL3 \cite{Lalazissis}, FRDM \cite{Moller97}, Gogny D1S \cite{AMEDEE} and DD-ME2 \cite{Lalazissis2005} are also presented for comparison.\\

As one can see from Fig.~\ref{S2n}, the two-neutron separation energies are getting smaller and smaller as the neutron number $N$ increases. Also, a sharp decline is clearly seen at $N=126$ both in the experimental and theoretical curves, which indicates the closed shell at this magic neutron number. Moreover, except for few lead isotopes, our calculated values are in a good agreement with the experimental data and have the lowest rms deviation in comparison with the results of NL3, Gogny D1S and DD-ME2, as we can see from Table~\ref{table4}.

\begin{table}[!htb]
	\caption{The root mean square deviations of the calculated two-neutron separation energies of $Pb$ isotopes from the experimental data, in comparison with other theoretical models.
		\label{table4}}
	\centering
	\begin{tabular}{lllll} 
		\hline\noalign{\smallskip}
		SLy4 	&   NL3		&  FRDM		&	Gogny D1S &   DD-ME2 \\ 
		\noalign{\smallskip}\hline\noalign{\smallskip}
		0.9679		&	1.1343	&  0.2394 	&	1.1948	  &   0.9847 \\  
		\noalign{\smallskip}\hline
	\end{tabular}
\end{table}

\subsection{Neutron, Proton and Charge radii}

\begin{figure}[!htb]
	\centerline{\psfig{file=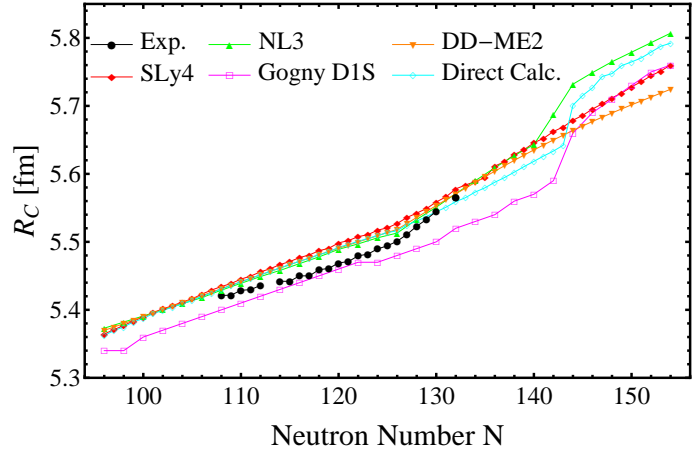,height=6cm}}
	\caption{(Color online) The calculated charge radii for $Pb$ isotopes obtained from the HFB calculations with Sly4, in comparison with experimental data and those obtained by FRDM, Gogny D1S, NL3,	DD-ME2, and Direct Calculation.}
	\label{Rc}
\end{figure}

The root-mean-square (rms) charge radius, $R_c$, is related to the rms proton radius, $R_p$, by :
\begin{equation}
R_c^2=R_p^2+0.64~(fm^2)
\label{eqRc}
\end{equation} 
where the factor $0.64$ in Eq.~(\ref{eqRc}) accounts for the finite-size effects of the proton. 
Fig.~\ref{Rc} shows the rms charge radii of lead isotopes calculated in this work. For the sake of comparison, it also shows the available experimental data \cite{Angeli2004} and the predictions of other nuclear models.

From Fig.~\ref{Rc}, Our calculated charge radii as well as those obtained in NL3 and DD-ME2 methods follow closely the empirical ones. The Gogny D1S calculations, on the other hand, underestimate the charge radii of Pb isotopes.

\begin{figure}[!htb]
	\centerline{\psfig{file=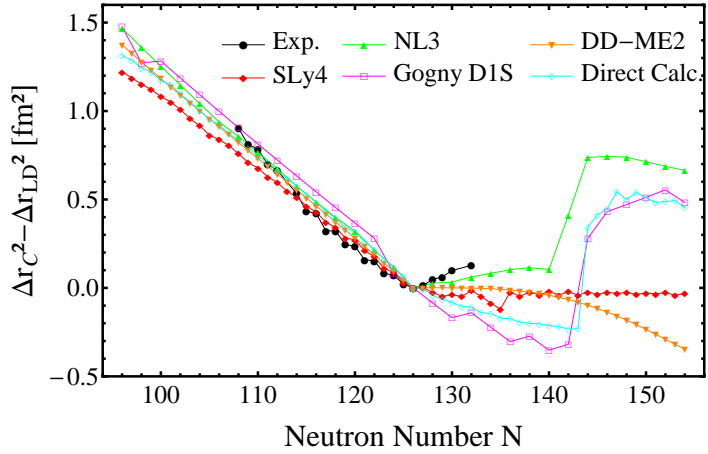,height=6cm}}
	\caption{(Color online) 
		The calculated isotope shifts ($\Delta r_c^2-\Delta r_{LD}^2$) for $Pb$ isotopes obtained from the HFB calculations with Sly4, in comparison with experimental data and those obtained by FRDM, Gogny D1S, NL3,
		DD-ME2, and Direct Calculation.}
	\label{Rc_shift}
\end{figure}

The calculated charge radii have been used to obtain the isotope shifts ($\Delta r_c^2=r_c^2(A)-r_c^2(208)$), where the nucleus $^{208}Pb$ has been taken as the reference point. 
In order to emphasize the deviation from the smooth trend, we have plotted in Fig. \ref{Rc_shift}
the modified isotope shifts ($\Delta r_c^2-\Delta r_{LD}^2$)
obtained by subtracting an equivalent of the liquid-drop difference $\Delta r_{LD}^2 = r_{LD}^2(A) -  r_{LD}^2(208)$. The mean-square radius of the liquid-drop model is given by $r_{LD}^2 =\frac{3}{5}r_0^2 A^{2/3}$ with $r_0 = 1.2$ fm \cite{Sharma1993,Tajima1993}.
Fig. \ref{Rc_shift} includes also the available experimental data \cite{Angeli2004} and the predictions of other nuclear models. All the data have been presented in the same fashion.

As we can see from Fig. \ref{Rc_shift}, the experimental data exhibit a conspicuous kink about $^{208}$Pb. Also, our calculated isotope shifts  reproduce this kink quit well. On the higher side of $^{208}$Pb, our results as well as the other theoretical models show a slight divergence from the experimental data and the slope of the theoretical values is also different from that of the experimental ones. This is due to the fact that most of the Skyrme forces have been fitted to the binding energies and charge radii of spherical nuclei, and therefore, they face an enormous problem in describing the isotopes shifts of nuclei away from $^{208}$Pb, as it has been demonstrated unambiguously in Ref. \refcite{Tajima1993}.\\

\begin{figure}[!htb]
	\centerline{\psfig{file=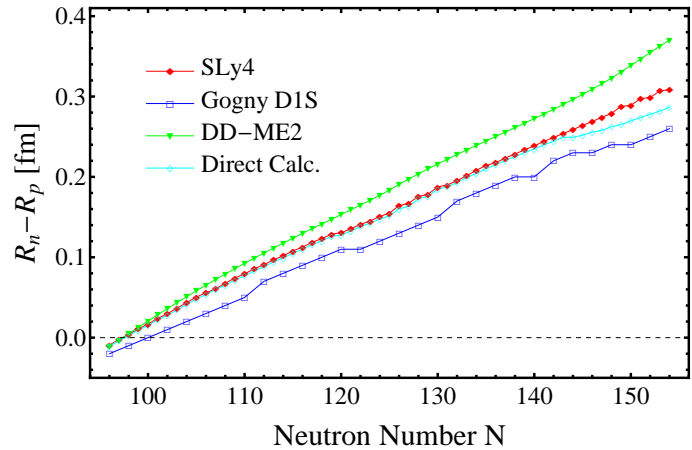,height=6cm}}
	\caption{(Color online) The calculated neutron skin thicknesses ($\Delta R=R_n-R_p$) for $Pb$ isotopes obtained from the HFB calculations with Sly4, in comparison with results obtained by Gogny D1S, DD-ME2, and Direct Calculation.}
	\label{R}
\end{figure}

Fig.~\ref{R} shows the difference between $R_n$ and $R_p$ ($\Delta R=R_n-R_p$), i.e. neutron skin, of lead isotopes obtained in our calculations as function of the neutron number $N$. The HFB calculations based on the D1S Gogny force \cite{AMEDEE} are shown for comparison as well as the results of the relativistic Hartree-Bogoliubov (RHB) model with the DD-ME2 effective interaction that we have calculated by using the computer code DIRHBZ \cite{code}. 

As it can be seen from Fig.~\ref{R}, In the vicinity of the
$\beta$-stability line
($N \approx Z$), the neutron and proton radii are approximately the same. The difference between $R_n$ and $R_p$ ($\Delta R=R_n-R_p$) increases monotonously with increasing the neutron number, in favor of developing a neutron skin. $\Delta R$ attains its maximum for $^{236}$Pb, in which it reaches $0.308$ fm in our calculation, and $0.370$ fm and $0.260$ fm in DD-ME2 and Gogny D1S nuclear models, respectively.

\subsection{Quadrupole deformation} 

The quadrupole deformation, $\beta2$, is also an important property  for describing the structure and shape of the nucleus. For certain nuclei, it can make a deformed shape more favored than the spherical one by increasing the nuclear binding energy.

\begin{figure}[!htb]
	\centerline{\psfig{file=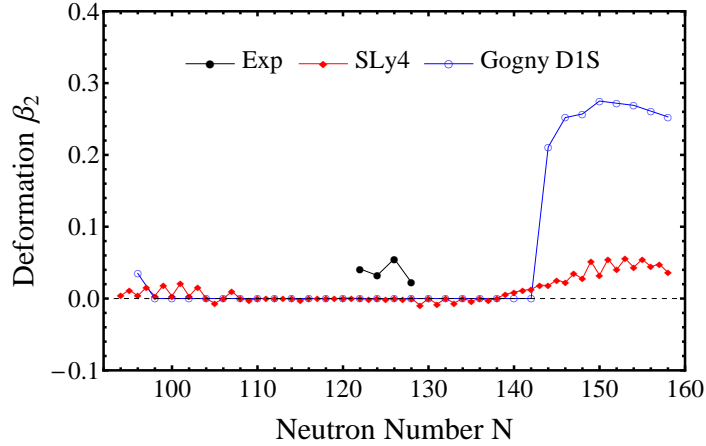, height=6cm}}
	\caption{(Color online) The calculated ground-state deformations, $\beta_2$, for $Pb$ isotopes obtained from the HFB calculations with Sly4, in comparison with experimental data and those obtained by Gogny D1S.}
	\label{beta}
\end{figure}

The quadrupole deformation values, $\beta_2$, for lead isotopes are plotted in Fig.~\ref{beta}. The results of our calculations are compared with both available experimental data \cite{beta_exp} and results of  HFB calculations based on the D1S Gogny force \cite{AMEDEE}.

As shown in Fig.~\ref{beta}, our results as well as those of Gogny D1S predict that all the isotopes from $N=94$ up to $N =142$ are almost spherical. For lead isotopes with neutron number higher than $N=142$, the prolate deformation increases and then saturates at a value close to $\beta_2=0.05$ in our calculation and $\beta_2=2.6$ in Gogny D1S results. 

The available experimental data show that the lead isotopes
from $N =122$ to $N =128$ exhibit weak deformation, which means that even the doubly closed shell nuclei and obviously spherical nuclei $^{208}Pb$ ($N=126$) exhibits "small deformations". This is due to the fact that the experimental $\beta2$ values are usually extracted from experimental $B(E2)\uparrow$ values (the reduced probability of the transition from the ground state of the nucleus to the first excited $2^+$ state) by using the Bohr model, which is valid only for well-deformed nuclei \cite{Robledo2012}. 

Here we should mention that the parameter set NL3 has difficulties to reproduce the proper deformations in light Pb and Hg isotopes \cite{Niksic2002}, and therefore a slightly modified set NL3* has been introduced by Lalazissis \textit{et al}. \cite{Lalazissis2009} by performing a new global fit of ground state properties of spherical nuclei and infinite nuclear matter.

\section{Conclusion}
\label{section4}
In the present work, we have studied the ground-state properties of even and odd lead isotopic chain, $^{178-236}$Pb, from the proton drip line to the neutron drip line in the framework of HFB theory with SLy4 Skyrme force. The calculations were made by using the computer code HFBTHO (v2.00d), in which
the pairing strength was given by our new generalized formula that gives, for each mass number A, the appropriate pairing strength $V_0^{n,p}$ for neutrons and protons instead of the default value which is constant regardless of the mass number. Our calculations reproduce the available experimental data very well including the nuclear binding energies, two-neutron separation energies. The parabolic behavior of the $BE/A$ has been  well reproduced in respect to the experimental curve. The calculated charge radii are also coherent with the experimental data for most nuclei. The neutron skin, in our calculation, reaches $0.308$  for $^{236}$Pb. Moreover, our results for quadrupole deformation gap reproduce the available experimental data quit well.

\section*{Acknowledgment}
The authors acknowledge valuable discussions with Dr. Nicolas Schunck from Lawrence Livermore National Laboratory.

\end{document}